\documentclass[prd,twocolumn,floatfix]{revtex4-2}

\newcommand{\omitfigs}{0}

\usepackage{amsmath}  % needed for \tfrac, \bmatrix, etc.
\usepackage{amsfonts}  % needed for bold Greek, Fraktur, and blackboard bold
\usepackage{graphicx}  % needed for figures
\usepackage{epstopdf}  % convert eps figures to pfd (taken care of automatically on modern systems)
\usepackage{amssymb}

\usepackage{epstopdf}

\usepackage{pgfplots}
\usetikzlibrary{pgfplots.groupplots}
\newcommand{\mytikz}[2]
{\ifodd \omitfigs
\else
\begin{tikzpicture}[scale=#1] \input{#2} \end{tikzpicture}
\fi
}

\newcommand{\be}{\begin{eqnarray}}
\newcommand{\ee}{\end{eqnarray}}

\newcommand{\brho}{ b}
\newcommand{\rmin}{r_{\!\rm min}}

\newcommand{\dt}{{\rm d}t}
\newcommand{\kB}{k_{\rm B}}

\newcommand{\dby}[2]{ \frac{{\rm d} #1}{{\rm d} #2}}

\newcommand{\ket}[1]{ \left| #1 \right> }

\newcommand{\nB}{ n_{\scalebox{.8}{$\scriptstyle \rm B$}} }

\begin{document}
\def \d {\rm d}

\ifodd 0
Highlights

Gravitational waves from nucleon elastic scattering are calculated.

The tensor contribution to the internucleon force allows gravitational quadrupole radiation even for S-wave scattering. 

This is the main source of a very-high frequency component to the 
stochastic background in the Solar System and in neutron stars.
\fi

\title{Gravitational bremsstrahlung from Yukawa and nucleon collisions}

\author{Andrew M. Steane}
\email{a.steane@physics.ox.ac.uk} % optional
%\ead{a.steane@physics.ox.ac.uk} 
%\affiliation{Department of Atomic and Laser Physics, Clarendon Laboratory, Parks Road, Oxford OX1 3PU, England.}
\address{Department of Atomic and Laser Physics, Clarendon Laboratory, Parks Road, Oxford OX1 3PU, England.}

\date{\today}

\newcommand{\lamC}{\lambdabar_{\rm C}}

\begin{abstract}
We obtain the gravitational emission from particles scattering via the
Yukawa interaction, presenting both classical and approximate quantum
results. We also estimate the contribution from the tensor part of the
internucleon interaction. 
This emission is the main source of a very-high frequency component to the 
stochastic background in the Solar System and in neutron stars.
The emission from the Sun (allowing for Debye screening) and from 
a typical neutron star are obtained. The gravitational wave luminosity
of the Sun is $41 \pm 10$ MW.
\end{abstract}

%\pacs{03.30.+p, 03.50.De,  04.20.-q, 04.40.Nr}
%PACS numbers: 03.30.+p, 03.50.De,  04.20.-q, 04.40.Nr, 04.40.-b

% 03.30.+p   = Special relativity
% 04.40.-b   = Self-gravitating systems; continuous media and classical fields in curved spacetime

% not 04.25.Nx, 04.30.Db

\maketitle

Gravitational bremsstrahlung has long been of interest from a fundamental
point of view. It has recently also attracted interest from an astrophysical
point
of view since it may be a non-negligible process in clusters of black holes
believed to be present in galactic nuclei. Previous work has focussed on
the $1/r$ potential, which applies to Coulomb collisions (such as those
between electrons in stellar interiors) and 
to hyperbolic encounters of stars or black holes moving
in a Newtonian gravitational potential. A survey is given
in \cite{Steane2024}.

When considering gravitational wave (GW) emission from collisions of
particles, the Yukawa potential is the next most important
case, because it gives a rough model of the internucleon potential
and because it is also relevant to collisions of charged particles 
in a plasma with Debye screening. The effect of Debye screening 
can be modelled to first approximation by replacing the Coulomb potential by
a Yukawa potential with a suitable decay length (the Debye length). 
The Yukawa potential was previously considered by Boccaletti \cite{Boccaletti1972}. In this paper we shall reproduce
Boccaletti's results and extend them in three respects. First (Section \ref{s.class}), we show
how the results of the calculation using classical physics can be
simplified. Secondly (Section \ref{s.quant}), we discuss the GW emission
after allowing, approximately, for a quantum treatment of the collision. 
One notable observation is that 
the classical calculation is inappropriate
when treating collisions of nucleons or electrons,
so the previous work was not in fact applicable to neutron stars or
stellar plasmas. Thirdly (Section \ref{s.tensor}), we
estimate the contribution from the tensor part of the internucleon
potential. The tensor part allows there to be 
an evolving quadrupole term in
the mass distribution even for a collision with spherically symmetric
initial conditions, such as an incident s-wave. 
In Section \ref{s.applied} the formulae are applied to calculating GW emission from neutron stars and from the Sun.

\section{Classical treatment} \label{s.class}

\subsection{Yukawa collisions}

\begin{figure}
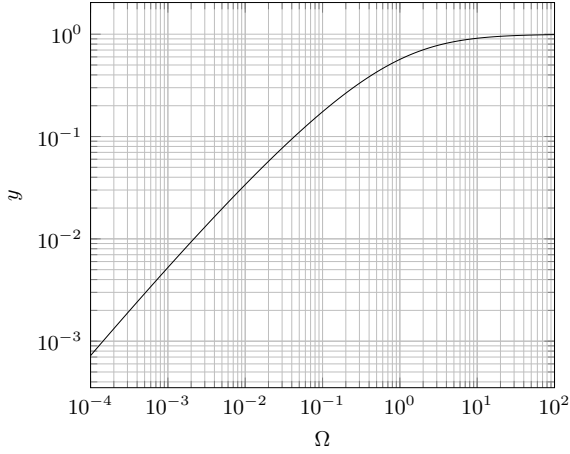

\mytikz{0.9}{yfig}
\caption{$y = r_{\rm min}/r_0$ as a function of $\Omega = E/V_0$. At high energy the colliding
	particles reach the region $\mu r \ll 1$ and then the minimum distance is $r_0$. At low
	energy the colliding particles approach to a distance of order $1/\mu$ so then $y \sim (\mu r_0)^{-1} = \Omega$.
	A more accurate approximation in the limit $\Omega \ll 1$ is $y \simeq \Omega \ln(1 + 1/\Omega)$.}
\label{f.yplot}
\end{figure}

The Yukawa potential is
\be
V(r) = \frac{V_0 e^{-\mu r}}{\mu r}
\ee
where $V_0$ and $\mu$ are constants ($\mu > 0$), and $r$ is the inter-particle separation. Let
\be
r_0 := \frac{V_0}{\mu E}
\ee
where $E = (1/2) m v_0^2$ is the collision energy. 
If $V_0 > 0$ then 
in the limit $|\mu r_0| \ll 1$, $r_0$ is the minimum distance achieved in a 
head-on collision. More generally, for a repulsive collision
($V_0 > 0$) that minimum distance is the solution of
\be
E = \frac{V_0}{\mu \rmin} \exp(-\mu \rmin).            \label{rmineq}
\ee
Let
$
y := {\rmin}/{r_0}.
$
Then we have that $y$ is the solution of
\be
y e^{y/\Omega} = 1     \label{yeq}
\ee
where
\be
\Omega := |\mu r_0|^{-1} = E / |V_0|.
\ee
The parameter $\Omega$ characterises the degree to which the distance of closest approach
differs from $r_0$. It is a useful parameter because it enables us to express solutions to
(\ref{rmineq}) in terms of given quantities and a function of a single quantity. That function (the solution to
(\ref{yeq})) is plotted in Fig. \ref{f.yplot}. We will show in the following that the 
parameter $\Omega$ is also useful in
characterising the degree to which GW scattering for Yukawa collisions differs from that
for Coulomb collisions. 

To calculate the gravitational wave (GW) emission, the classical treatment 
proceeds just as for the Coulomb case
described in \cite{Carmeli1967,Dehnen1985,Steane2024} except that now one finds
\be
\dddot{D}_{ik} \dddot{D}^{ik} = 
\frac{24 V_0^2}{\mu^2 r^4} e^{-2 \mu r}
\left( \zeta v^2 + \xi v_\perp^2 \right)
\ee
where $D_{ij}$ is the quadrupole tensor, and
\be
\zeta &=& (1 + \mu r - \mu^2 r^2)^2, \nonumber \\
\xi &=& 11 + 22 \mu r + 13 \mu^2 r^2 + 2 \mu^3 r^3 - \mu^4 r^4.
\ee
Hence the emitted power at any given moment during a collision is
\be
L_{\rm GW} = \frac{8 G}{15 c^5} V_0^2 e^{-2 \mu r} \frac{1}{\mu^2 r^4}
\left( \zeta v^2 + \xi v_\perp^2 \right).
\ee

We can express the total power generated in some small volume $V$ of
a gas or a plasma, as a result of collisions between particles of types 1 and 2, as
\be
P = V n_1 n_2 \left< v_0 \Sigma \right>          \label{PSigma}
\ee
where $n_i$ are the number densities of the particles and
\be
\Sigma  &=& \int_{-\infty}^\infty \dt
\int_0^\infty 2 \pi \brho  \, \d\brho   \, L_{\rm GW}   \nonumber\\
&=& 2 \, \int_{r_0}^\infty \frac{\d r}{|\dot{r}|} 
\int_0^{\brho  _{\rm max}} 2 \pi \brho  \, \d\brho   \, L_{\rm GW} 
\label{intclass}
\ee
where $\brho$ is the impact parameter.
The integration with respect to $\brho$
is described in \cite{Steane2024}. We obtain 
\be
\Sigma = \frac{32\pi G}{9 c^5} \frac{V_0}{\mu} 
m v_0^3  \chi  \label{chidef}
\ee
where
\be
\chi = \frac{r_0}{10} \int_{r_{\rm min}}^\infty \d r \,
e^{-2\mu r}
\left( \frac{3\zeta + 2\xi}{r^2} \right)
 \left(1 - \frac{r_0}{r} e^{-\mu r}\right)^{3/2} \!.
\label{chiYuk}
\ee 
This equation was first obtained by Boccaletti
(there are minor errors in Boccaletti's equations (10) and (11) but
his equation (12) is correct up to the sign of $V_0$;  his
symbol $\chi$ corresponds to our $\Sigma$).

\begin{figure}
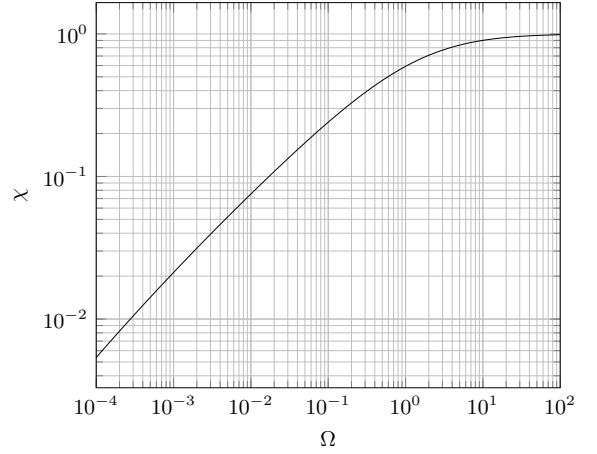

	\mytikz{0.9}{chi0fig}
	\caption{$\chi$ (eqn (\ref{chix})) as a function of $\Omega = E/V_0$. At high energy the
		value tends to 1, which is the result for a Coulomb collision with an interaction
		strength $V_0/\mu$. At low energy
		$\chi$ is smaller, since it reflects the effect of the exponential term in the Yukawa
		potential which reduces the interaction strength at large distance. At $\Omega \ll 1$
		we have $\chi \simeq \Omega^{3/4} \ln ( 1 + \Omega^{-3/4}).$}
	\label{f.chivsOmega}
\end{figure}

Now let us introduce a change of variable in the integral in (\ref{chiYuk}). Let
$
x := {r}/{r_{\rm min}} = {r}/({r_0 y}).
$
Then we have $\mu r = y x /\Omega$, and for $V_0 > 0$ one obtains
\be
\chi = \frac{1}{10 y} \int_1^\infty \!\!{\d} x \, 
e^{-2 y x / \Omega}
 \frac{(3\zeta + 2\xi)}{x^2}
\left[1 - \frac{ e^{-y x/\Omega} }{y x}\right]^{3/2} \!\!.   
     \label{chix}
\ee
In this equation, the functions $\zeta$ and $\xi$ are functions of $\mu r =  y x/ \Omega$,
hence the integral is a function of $\Omega$ and $y$ alone. But $y$ is a
function of $\Omega$ alone, so we deduce that $\chi$ is a function of $\Omega$ alone.
One finds $\chi \rightarrow 1$ in the limit $\Omega \rightarrow \infty$ and
$\chi$ falls below 1 when $\Omega \sim 1$. The overall dependence is shown
in Fig. \ref{f.chivsOmega}. This observation is the first main result of
the present work: the GW emission is expressed
in (\ref{chidef}) in terms of the given parameters, such that
the function $\chi$ is a function of a single parameter.

\section{Estimating the quantum effects}  \label{s.quant}

So far we adopted an entirely classical treatment and obtained precise
formulae. In this section we consider GW emission for Yukawa collisions when 
quantum effects are significant. We estimate 
the cross section $\Sigma$ in this case with a precision to be discussed.
The methods are most accurate for small $\eta$ (defined in (\ref{etadef})) and large $\Omega$.

In order to find the regime where the quantum description of the
trajectory differs significantly from the classical one, we define
the {\em Born parameter} by 
\be
\nB := \frac{\rmin}{2 \lambdabar_{\rm dB}} 
= \frac{m v \rmin}{2 \hbar}
\ee
where $\lambdabar_{\rm dB} = \hbar / m v$. 
$\nB \gg 1$ is the classical limit. 
It will be useful
also to describe the potential in terms of dimensionless parameters
$\{\eta,\; A\}$ defined by
\be
\eta := \frac{\mu \hbar}{m c}, \qquad A := \frac{V_0}{\hbar c \mu} \;=\; \frac{V_0}{m c^2 \eta}.
\label{etadef}
\ee
Values of $\eta$ and $A$ suitable for collisions of nucleons and
for the treatment of electrons in the Sun are shown in table~\ref{t.par}.

\begin{table}
\begin{tabular}{lcc}
\hline
    & $A$ & $\eta$ \\
\hline
%nn  & $-0.0758$  & $0.294$ \\
nn  & $-0.0273$  & $0.042$ \\
ee  & $0.00730$  & $0.035$ \\
ep  & $-0.00730$ & $0.017$ \\
eHe$^{++}$ & $-0.0146$  & $0.017$ \\
\hline
\end{tabular} 
\caption{Values of Yukawa potential parameters for various collision partners.
The values for nn are chosen so as to fit (approximately) the empirical
elastic cross-section reported in \cite{Brown2006}. The values for collisions involving electrons indicate the effect of Debye screening in the core of the Sun.}
\label{t.par}
\end{table}

\subsection{Repulsive collisions}

\begin{figure}
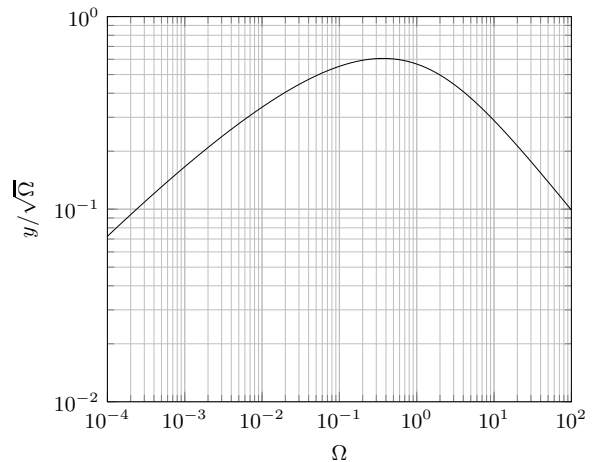

\mytikz{0.9}{yoverfig}
\caption{$y/\sqrt{\Omega}$ as a function of $\Omega = E/V_0$. 
For a repulsive Yukawa collision this
can be used to obtain the Born parameter as a function of collision energy;
see eqn \ref{nBy}).}
\label{f.yover}
\end{figure}

In the repulsive case, 
from the above definitions one obtains
\be
\nB = \sqrt{ \frac{A}{2 \eta} } \frac{y}{\sqrt{\Omega}}.
\label{nBy}
\ee
The quantity $y / \sqrt{\Omega}$ is plotted in figure \ref{f.yover}.
It has a maximum value of about $0.6$. 
The collisions we shall be treating have $A < \eta$ so we deduce that
the Born parameter will always be less than 1. In this respect
Yukawa collisions differ from Coulomb collisions. In the Coulomb case
there is a classical limit (high Born parameter) at low collision energy.
For Yukawa collisions the exponential term in the potential ensures that, for typical parameter values, the colliding partners can approach closely even
for low collision energy, with the result that $\nB$ does not exceed~$1$.

One expects the GW scattering to fall below the classical prediction
when $\nB < 1$. This is because the spread of the wavefunction tends
to smooth out the mass distribution, resulting in a lower rate of
change of the quadrupole moment.
The contribution of soft gravitons to the GW scattering is given by
\be
\Sigma_{< \Lambda} \simeq  \frac{8 G}{5\pi c^5} m^2 v^4  \frac{\Lambda}{\hbar}
\tilde{\sigma}
\label{Sigsoft}
\ee
where
\be
\tilde{\sigma} = \int \dby{\sigma}{\Omega} \sin^2 \theta \, \d\Omega \, .
\ee
This is Weinberg's {\em soft photon theorem}. \cite{Weinberg1965,Carmeli1967,Weinberg2019} The particular form given here
is discussed in \cite{Steane2024}.
In this context, {\em soft} gravitons are those which have 
negligible impact on the energy-momentum for lines in Feynman diagrams
that are near the mass shell (such as external lines). 
$\Lambda$ is an energy cut-off discussed below
 and $\d\sigma/\d\Omega$ is the differential cross
section for the collision in the absence of radiant emission. 

The soft photon theorem accurately gives the contribution of soft photons
to the spectrum at frequencies up to $\Lambda / \hbar$ if $\Lambda$ is
taken small enough, which means small compared to kinetic energies in
the collision. In the repulsive case
one can also use (\ref{Sigsoft}) to roughly estimate the
total scattering by taking a suitable value of $\Lambda$. That value is
of the order of $E$ since the spectrum extends up to $E$, but one
would expect that to take $\Lambda = E$ would overestimate 
the total $\chi$, and $\Lambda = E/4$ would underestimate. Thus on
general grounds one may propose $\Lambda = (0.5 \pm 0.25) E$, thus
giving theoretical uncertainty of about a factor 2. In order to improve
on this we employ the known results for
the Coulomb potential to provide a calibration \cite{Galtsov1974,Galtsov1976,Steane2024}.
The soft photon theorem cannot be used directly for the Coulomb potential
because the cross-section $\tilde{\sigma}$ then diverges. Therefore we
proceed in two steps. First we calibrate an integral described below 
(eqn (\ref{chixrmod})) which
can be used in the limit $\eta \rightarrow 0$ as well as for
$\eta$ up to $0.1$, then we adjust $\Lambda$
so as to match this integral at $\eta = 0.035$. By this procedure we
obtain $\Lambda \simeq 0.7 E$. We expect the absolute value of $\chi$
obtained this way to be accurate to about a factor 2 for $0.01 < \eta < 0.1$.
Relative values will be more accurate.

The quantity $\tilde{\sigma}$ scales with $V_0$ in the same way as
the total elastic collision cross section in the absence of GW emission,
$\tilde{\sigma} \propto (V_0/\mu)^2$. Therefore the soft photon theorem
predicts $\Sigma \propto (V_0/\mu)^2$ whereas the classical calculation
gives $\Sigma \propto (V_0/\mu)$. At first sight this may appear to be an
inconsistency, but it is not. It is simply that the scaling changes in the quantum limit. A similar scaling is found for the
repulsive Coulomb potential in the quantum limit, where the quantum 
(Born approximation) calculation gives  
$\Sigma = (9/2\pi) \nB \Sigma_{\rm classical}$ and the Born parameter
is proportional to the interaction strength \cite{Steane2024}.

\begin{figure}
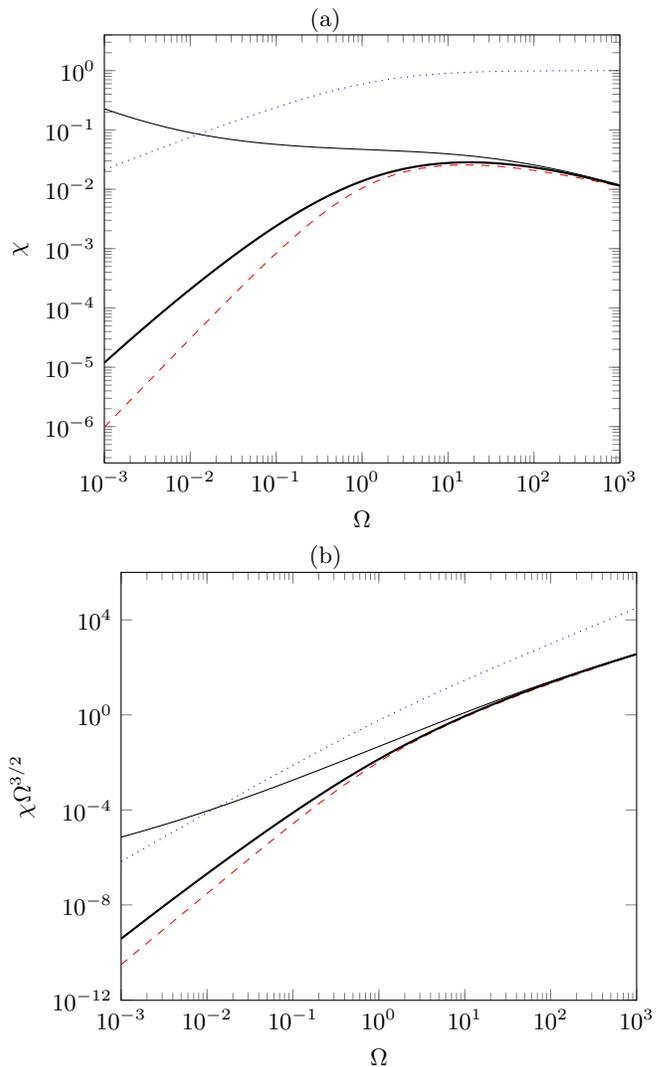

(a)
	\mytikz{1}{chisfig}
(b)	\mytikz{1}{Sigmafig}
\caption{
(a): estimates of $\chi$ allowing for quantum effects,
as a function of $\Omega = E/V_0$, for parameter values $\{|A|,\,\eta\} = \{0.007297, 0.0346\}$.
Dashed red line: soft photon theorem, repulsive case: eqn (\ref{chis}), 
taking $\Lambda = 0.7 \, E$.
Thick black line: modified classical integral (\ref{chixrmod}), repulsive case.
Thin black line: attractive case, modified classical integral (\ref{chixadj}).
The dotted blue line shows the classical result for the
repulsive case (eqn \ref{chix}), for comparison.
(b): corresponding graphs of $\chi \Omega^{3/2}$. The cross-section
$\Sigma$ is proportional to this (c.f. (\ref{chidef})).}
	\label{f.chisoft}
\end{figure}

By substituting (\ref{tildesigYuk}) into (\ref{Sigsoft}) and
defining $\chi$ as in (\ref{chidef}) we obtain
\be
\chi_{< \Lambda} \simeq 
 \frac{9}{5 \pi \sqrt{2}} A \frac{\Lambda}{E} \sqrt{ \frac{ m c^2}{E} }
\left[ \left(\frac{1}{2} + \frac{1}{z} \right) \ln (1+z) -1 \right]
\label{chis}
\ee
where 
\be
z = 4 k^2 / \mu^2 = 8 E / mc^2 \eta^2 = 8 \Omega A / \eta.
\ee
The result is shown in Fig. \ref{f.chisoft} for example parameter
values, where it is compared with the classical result (\ref{chix}).
We find that the quantum calculation gives a value lower than
the classical result by a factor of the order of the Born parameter.
In this respect the difference between quantum and classical 
is similar to that for Coulomb collisions. 

\begin{figure}
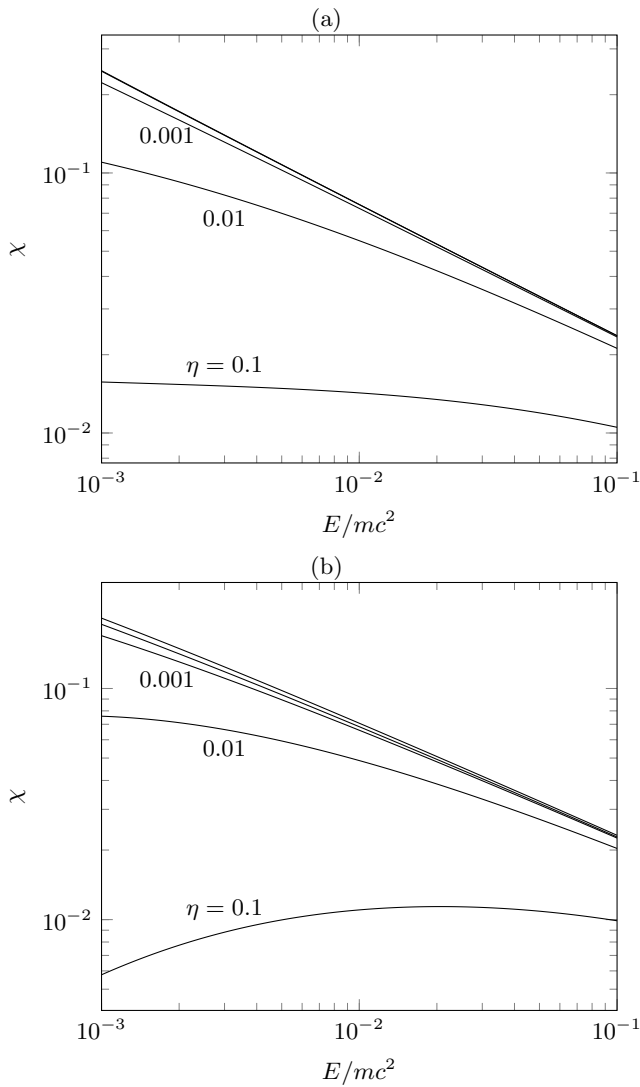

(a) \mytikz{1}{chivE1}
(b)	\mytikz{1}{chivE2}
\caption{$\chi$ as a function of $E/mc^2$ for 5 values of $\eta$,
for $|A| = \alpha = 0.007297$. (a) attractive case, 
(\ref{chixadj}); (b) repulsive case 
(\ref{chixrmod}).
The $\eta$ values are $\{0, 0.0001,0.001,0.01,0.1\}$. $\chi$
decreases with $\eta$ at any given $E$. The curves for $\eta=0$
were obtained from equations (33)--(36) of \cite{Steane2024}.
}
	\label{f.etalimit}
\end{figure}

An alternative way to estimate the GW scattering is to take
the classical integral (\ref{chiYuk}) and modify it so as to reduce
the contribution from the region at $r < \lambda_{\rm dB}$
where $\lambda_{\rm dB}$ is the de Broglie wavelength.
It was shown in \cite{Steane2024} that this gives good accuracy for
repulsive Coulomb collisions if one simply
shifts the lower limit by about half the 
de Broglie wavelength associated with the initial speed $v_0$. 
For the Yukawa potential we adopt a similar method, but with
a smooth cut-off, which is achieved by multiplying the integrand 
in (\ref{chiYuk})
by the filter function $r^2 / (r^2 + \lambda^2)$ with
$\lambda$ a parameter of the order of the de Broglie wavelength.
We have
\be
\lambda_{\rm dB} = \frac{2\pi\hbar}{\sqrt{2 m E}} = 2\pi r_0 \sqrt{\frac{\eta \Omega}{2A}}         \label{lamOm}
\ee
so the modified integral is
\be
\chi &=& \frac{1}{10 y} \int_1^\infty {\d} x \, e^{-2 y x / \Omega}
\left( \frac{3\zeta + 2\xi}{x^2 + \beta 2 \pi^2\eta \Omega/(A y^2) } \right)
\nonumber\\
&& \rule{10ex}{0pt} \times 
\left(1 - \frac{1}{y x} e^{-y x/\Omega}\right)^{3/2}       \label{chixrmod}
\ee
where $\beta$ is adjusted so as to make the result agree with the known
behaviour in the limit $\eta \rightarrow 0$ (i.e. the Coulomb potential).
Figure \ref{f.etalimit} shows the behaviour for low $\eta$. The Coulomb
limit is obtained correctly by taking $\beta = 3/4$.
Figure \ref{f.chisoft} shows a comparison between this result and that
of the soft photon theorem. They agree within a factor 2
over a wide range of $\Omega$, though not at very small $\Omega$.
This rough agreement also holds for a wide range of values of the parameters
$A$ and $\eta$ but not for extreme values. The main thing to note is that
the above treatment lends support to the view that
the soft photon theorem estimates the total GW scattering in the 
repulsive case when $\Lambda \simeq 0.7 \, E$.

\subsection{Attractive collisions}

In the case of attractive collisions the soft photon theorem cannot
be used to get a good estimate of the total scattering. This is because
one has no way to set the cut-off $\Lambda$ so as to extrapolate the
soft emission in order to estimate the total emission. We therefore
employ a rough estimate obtained by modifying the classical integral.

In the attractive case we have $r_{\rm min} = 0$ (classically) 
and there is no need
to introduce $y$. The classical calculation yields
\be
\chi = \frac{1}{10} \int_0^\infty {\d} x \, e^{-2 x / \Omega}
 \frac{(3\zeta + 2\xi)}{x^2} 
\left(1 + \frac{1}{x} e^{-x/\Omega}\right)^{3/2}        
\ee
where now $x = r / |r_0|$. This integral diverges. The quantum result
avoids the divergence, which can be understood from the rough physical
picture of replacing classical point particles with wavepackets of finite
position-spread. The main consequence of this spreading is to reduce
the contribution to the integral at low $x$. We have not calculated this
reduction. For a rough estimate we simply multiply the
integrand by a filter of the form of the form $r^2 / (r^2 + \lambda^2)$
as described above for the repulsive case, and similarly adjust
the second divergent term, resulting in the following form:
\be
\chi &\simeq& \frac{1}{10} \int_0^\infty \!{\d} x \, e^{-2 x / \Omega}
\left( \frac{3\zeta + 2\xi}{x^2 + 3\pi^2\eta\Omega/2A} \right)
\nonumber\\
&& \rule{4ex}{0pt} \times
\left(1 + \frac{1}{\sqrt{x^2 + 3\pi^2\eta\Omega/2A}} e^{-x/\Omega}\right)^{3/2} \!\!.
    \label{chixadj}
\ee
The result is shown in figure \ref{f.chisoft}. This estimate 
is most accurate for $\eta < 0.1$ and $\Omega > 1$. For smaller
$\Omega$ it is less accurate but serves
to convey the general impression of the dependence: we find a monotonic decreasing function of $\Omega$, 
which approaches the value of the repulsive case at large $\Omega$.

The above formulae are applied to calculations of GW emission in neutron stars and the Sun in section \ref{s.applied}.

\section{Tensor interaction toy model}  \label{s.tensor}

It is well known that the internucleon interaction contains tensor
as well as scalar contributions.
This leads, for example, to the
electric quadrupole moment of the deuteron. We should like to estimate
the contribution of the tensor term to GW emission in low-energy
collisions. Here `low energy' signifies non-relativistic, which for
nucleons means energy below some tens of MeV
(the reduced mass for a nn or np or pp collision is 469 MeV).

We shall consider the case of an incident s-wave (i.e. zero orbital
angular momentum in the incident wave).
By conservation of angular momentum, s-wave scattering on a central potential
would give no GW emission. 
Therefore in this case the GW emission is wholly associated
with the tensor part of the internucleon interaction. 

For present purposes it will be sufficient to consider 
a very simple quantum-mechanical model of a pair of
nucleons undergoing a collision. This toy model describes the system in
terms of two discrete energy eigenstates. These states
are a rough stand-in for two regions of the continuum, such that region $i$
extends from energy $E_i - \delta_i$ to $E_i + \delta_i$ with $\delta_i$ determined by
$\int_{E_i - \delta_i}^{E_i + \delta_i} \rho(E) \d E = 1$ where $\rho(E)$ is
the density of states and we pick $E_2 = E_1 + \delta_1 + \delta_2$.
In order to motivate this simple approach a little, note that
since the deuteron is bound, and nuclear forces are charge-symmetric
to first approximation, it follows that the diproton and dineutron are 
almost bound; that is,
there would be a bound state if the potential well were a little deeper or wider.
This implies that there is a resonance or quasi-bound state in the
continuum at low energy. 

We adopt
\be 
\left< \dddot{Q}_{ij} \dddot{Q}^{ij} \right> \simeq
\left( \frac{\d^3}{\dt^3}\langle \psi | \hat{Q} | \psi \rangle \right)^2
\label{QQapp}
\ee
where $\hat{Q}$ is the (mass) quadrupole moment operator and
$\psi$ is a normalized wavefunction describing a pair of nucleons
moving under the nuclear (strong) force with initial condition an $s$-wave
state with spatial extent of order fm. The form of
$\psi$ and its evolution over time have now to be considered. 
We consider np scattering in the first instance.

Consider a system like the deuteron but with two bound states.
The idea is to use this to estimate what rate of change of 
quadrupole moment is liable to arise in an elastic collision governed by 
the nuclear force. Our toy system has symmetry requirements 
(isospin singlet, spin triplet) such that the initial spatial state has
even parity. 
Selection rules for quadrupole radiation are obeyed if the final state, 
after emission of a graviton, is $^3$D or $^1$P (for an initial state $^3$S).
The system has a ground state
$\ket{g} = a \ket{S} + b \ket{D}$ and an excited state
$\ket{e} = b^*\ket{S} - a^* \ket{D}$ where $\ket{S}$ and $\ket{D}$
have orbital angular momentum $0$ and $2$ respectively;
$a$ and $b$ are constants to be determined ($|a|^2 + |b|^2 = 1$).
The system is prepared at $t=0$ in the state $\ket{\psi(0)} = \ket{S} = a^* \ket{g}
+ b \ket{e}$, hence $\ket{\psi(t)} = \exp(-i E_g t/\hbar) \left( a^* \ket{S}
+ b \exp(-i\omega t) \ket{e} \right)$ where $\omega = (E_e - E_g)/\hbar$.
If $a$ and $b$ are real then for this system we find
\be
\frac{\d^3}{\dt^3} \langle \psi | \hat{Q} | \psi \rangle 
=  2  a b \omega^3 \sin(\omega t) Q_{ge}    \label{d3app}
\ee
where $Q_{ge} = \langle g | \hat{Q} | e \rangle = (b^2 - a^2) Q_{SD}
- ab Q_{DD}$ assuming the quadrupole matrix elements are real.
We have also $Q_{gg} = 2 a b Q_{SD} + b^2 Q_{DD}$. For the deuteron
one finds $|Q_{DD}| \ll |Q_{SD}|$ so we can assume this here,
and therefore
\be
Q_{ge} \simeq - (a^2 - b^2)  Q_{gg} / (2 ab).
\label{Qge}
\ee
For the deuteron one has $b \simeq 0.2$ and the electric quadrupole
moment is $(e/3)(0.00288 (2)\,{\rm barn})$ \cite{Krane1988}.
For the present calculation
a reasonable estimate is therefore $b = 0.2$, $a=0.98$ and $Q_{gg} = 
(2m_{\rm p}/3)(0.003 \,{\rm barn})$. This is not precise: the mass
distribution need not, and presumably does not, follow the charge 
distribution exactly.

We now claim that (\ref{d3app}) can be adopted as an approximation 
in (\ref{QQapp}) if we replace $\sin(\omega t)$ by a constant of order 1
and adopt a value of $\omega$ corresponding to the timescale of the collision. 
This is reasonable because the deuteron wavefunction extends well outside
the range of the potential well, so the evolution of the wavefunction in the 
elastic collision has similar distance- and time-scale to those of the toy model, 
and, crucially, a similar tensor component in the potential which
drives the development of a quadrupole moment from an s-wave initial condition.
To estimate $\omega$ we note that the mean kinetic energy in the ground state
of the deuteron is around 23 MeV and the tensor interaction (which is the part
causing the quadrupole) is of this same order.
We shall adopt
$\hbar \omega \simeq (20 \pm 10)\,$MeV. (This value of $\omega$ is also the inverse of the
time it takes a classical proton of kinetic energy $20\,$MeV to travel 2 fm).
We obtain for this contribution to the gw scattering,
\be
\Sigma_{\rm s1} \approx  (\pi G/10 c^5)  \sigma_{\rm s1} \omega^5 Q_{gg}^2 
\label{Sigs}
\ee
where the subscript s1 stands for `s-wave, spin 1' and $\sigma_{\rm s1}$ is the triplet $s$-wave
contribution to the collision cross section. For example at $\sigma_{\rm s1} = 
0.1$ barn we obtain $\Sigma_{\rm s1} \approx 10^{-38}\,$eV barn.
The uncertainty 
of this estimate is mainly in $\omega$ and $Q$. If both
are uncertain by a factor $2$ then the result is uncertain by a factor 32.

The case of nn and pp collisions differs from the above in one main respect: now the isospin state is symmetric so the spatial/spin part is antisymmetric. The tensor interaction is zero for the singlet, which implies that at low angular momentum we only get a dynamic quadrupole moment in the triplet states,
so we have to consider $^3$P and $^3$F.
This implies that in (\ref{Sigs}) $\sigma_{\rm s1}$ should be replaced by $\sigma_{\rm p1}$. $Q_{gg}$ and $\omega$ will also change, but one
would not expect them to change by a large factor. 
For low-energy pp collisions the Coulomb barrier must also be taken
into account.

The formula (\ref{Sigs}) predicts that for np scattering this contribution
to the GW emission is similar to the contribution from scattering off
the scalar potential (\ref{chis})
for a collision energy around 3 MeV, and at lower collision energies it dominates.

\section{Applications}  \label{s.applied}

\subsection{Neutron star}

We estimate gravitational bremsstrahlung from collisions of neutrons 
in a neutron star during the earlier part of the star's evolution when
the temperature is high. Following Boccaletti, the total emitted power
is roughly
\be
P \simeq V n^2 v_0 \Sigma (\kB T / E_{\rm F})^3
\ee
where $V$ is the volume, $n$ the neutron number density, $v_0$ is the mean speed
at the Fermi energy and $E_{\rm F}$ is the Fermi energy. 
The factors of $(\kB T/E_{\rm F})$ account for the fact that in the Fermi-Dirac distribution about $n \kB T /E_{\rm F}$ of the neutrons are available for scattering, and a further factor comes from the reduction of the cross-section 
owing to the reduction in the density of final states. 

As a typical case we adopt
the parameters $V = 10^{13}\,{\rm m}^3$, $n = 6 \times 10^{43}\,{\rm m}^{-3}$
(thus 1 solar mass), $v_0 = 10^8\,$m/s, $T = 10^9\,$K, $E_{\rm F} = 30\,$MeV.
Since neutron collisions are attractive, to calculate $\Sigma$ 
we must use the quantum estimate for attractive collisions, which is
similar to the result of the soft photon theorem at these energies.
We find $\chi \simeq 0.07$ and $\Sigma \simeq 3.5 \times 10^{-37}\,$eV barn.
This gives $P = 5 \times 10^{17}\,$watt. This is about two orders of magnitude
smaller than Boccaletti's estimate for the same kind of star \cite{Boccaletti1972}.
One order of magnitude discrepancy is accounted for by the fact that
Boccaletti used a
repulsive potential and a classical integral, both of which are incorrect.

Although the neutron stars are, presumably, among the brightest sources
of gravitational waves at these high frequencies ($10^{22}$ Hz), the 
result is still rather dim:
for the neutron star associated with the Crab nebula the flux on Earth
is about 1 graviton per km$^2$ per day.
It would be interesting, nonetheless, to investigate whether there is any observable effect of the gravitational noise inside and near a neutron star.

\subsection{The Sun}

Gravitational bremsstrahlung in the Sun arises mainly from
collisions among electrons, protons and $^4$He nuclei. Previously
the effects were calculated for motion under the Coulomb potential \cite{Steane2024}.
However there is substantial Debye screening in the Sun: in the solar core
the Debye length is about 22 pm. Elements of a suitable kinetic theory
were developed by Gal'tsov and Grats but they did not estimate the net effect \cite{Galtsov1983}. Here we estimate the effect of
Debye screening by modelling the collisions using the Yukawa potential.

When $\chi$ has little dependence on $\Omega$, we have $v \Sigma \propto E^2$
so we can perform the velocity average in (\ref{PSigma}) by taking $\Sigma$
at the r.m.s. energy in the thermal distribution. This is $E = 2 \kB T
\simeq 2.7\,$keV in the solar core. The parameters of the screened potential
are $A = \alpha$, $\mu = 1/\lambda_{\rm D}$ hence $V_0 = \alpha \hbar c \mu \simeq 65\,$eV, $\eta = 0.035$ for ee collisions. This yields $\Omega \simeq 42$
at the r.m.s. energy. The classical result for repulsive collisions, (\ref{chix}) and Fig. \ref{f.chivsOmega}, gives $\chi = 0.97$ at this
value of $\Omega$
and therefore almost no change from the prediction for Coulomb collisions
with no screening. However, in the solar core the Born parameter is
about $0.06$ so we must use the quantum calculation. By comparing 
(\ref{chixrmod}) and (\ref{chixadj}) with the Coulomb case we
find the GW emission from ee scattering is reduced by a factor $0.45$
and that for ep and eHe scattering is reduced by a factor $0.58$,
compared to the unscreened result, after integrating over the whole
Sun using the standard solar model.
We thus find that the total solar output is
$P \simeq 41 \pm 10\,$MW. The uncertainty in this result 
is mainly from the imprecision attached to equations 
(\ref{chixrmod}) and (\ref{chixadj}).

It is interesting to explore whether the pp collisions also
give a significant contribution via the tensor interaction of
the nuclear force. The cross section is reduced by the Coulomb
barrier. For the s-wave term,  
a standard calculation in the WKB approximation gives
\be
\frac{r_0^2 | \psi(r_0) |^2}{r_E^2 |\psi(r_E)|^2} \simeq
 e^{-\sqrt{E_{\rm C}/E} }
\ee
where $\psi(r)$ is the s-wave contribution to the wavefunction,
$r_0$ is some small distance of order fm, 
$r_E$ is the classical turning point
for a head-on collision at energy $E$ 
and $E_{\rm C} \equiv 2 m c^2 (\pi \alpha Z_1 Z_2)^2$.
We have $E_{\rm C} = 493$ keV for a pair of protons
(using the reduced mass $m = m_{\rm p}/2$). 
The s-wave cross-section for processes involving close collisions at distances
of order fm is reduced by the above exponential factor, which
evaluates to $1.4 \times 10^{-6}$ at $E=2.7\,$keV. The p-wave
cross-section is reduced still further. 
If it were not for this, the contribution to GW emission
from the Sun owing to
pp scattering via the strong force would be non-negligible.

\appendix
\section{Calculation of $\tilde{\sigma}$}  \label{s.cross}

We adopt the standard treatment of a collision, where a particle of
mass $m$ (i.e. the reduced mass) scatters off a fixed collision centre.
Let $k$ be the wavevector, such that the 
collision energy is $E = (\hbar k)^2/(2m)$. 
For the Yukawa potential in first order Born approximation,
one finds the differential scattering cross-section
\be
\dby{\sigma}{\Omega} = \left| f^{(1)} \right|^2 = 
%4 \mu^2 \lambda^2 \frac{1}{(\mu^2 + q^2)^2} 
\frac{4 A^2}{\lamC^2} \frac{1}{(\mu^2 + q^2)^2} 
\ee
%where $\lambda = A / \eta$ and
where $\lamC = \hbar / mc$ is the reduced Compton wavelength and
$q = 2k \sin(\theta/2)$ (thus $\hbar q$ is the momentum exchange).
This yields the ordinary collision
cross-section $\sigma = 16 \pi (A/\eta)^2 (\mu^2 + 4 k^2)^{-1}$ and
\be
\tilde{\sigma} 
= 8\pi \frac{A^2}{\lamC^2 k^4} 
\left[
\ln \left( \frac{\mu^2 + 4 k^2}{\mu^2} \right) \left( \frac{\mu^2}{4k^2}
+ \frac{1}{2} \right) - 1 \right] \!.
\label{tildesigYuk}
\ee
This tends to $(32 \pi/3) \lamC^2 A^2/\eta^4$ at $k \ll \mu$ and 
$8\pi A^2 \lamC^{-2} k^{-4} \ln (2k/\mu)$ at $k \gg \mu$.

\bibliography{gravitycosmology}

\end{document}